\documentclass{iau}
\usepackage{graphicx}

\title[Gas accretion from halos to disks] 
{Gas accretion from halos to disks: observations, curiosities, and problems
}

\author[Bruce G. Elmegreen]   
{Bruce G. Elmegreen$^1$
 }
\affiliation{$^1$IBM T.J. Watson Research Center, 1101 Kitchawan Road, Yorktown Heights, NY 10598, USA \\
email: {\tt bge@us.ibm.com} \\[\affilskip]}

\pubyear{2015}
\volume{317}  
\jname{The General Assembly of Galaxy Halos: Structure, Origin and Evolution}
\editors{Bragaglia, Arnaboldi, Rejkuba, Romano: eds.}
\begin{document}

\maketitle

\begin{abstract}
Accretion of gas from the cosmic web to galaxy halos and ultimately their disks is
a prediction of modern cosmological models but is rarely observed directly or at
the full rate expected from star formation. Here we illustrate possible
large-scale cosmic HI accretion onto the nearby dwarf starburst galaxy IC10,
observed with the VLA and GBT. We also suggest that cosmic accretion is the origin
of sharp metallicity drops in the starburst regions of other dwarf galaxies, as
observed with the 10-m GTC. Finally, we question the importance of cosmic
accretion in normal dwarf irregulars, for which a recent study of their far-outer
regions sees no need for, or evidence of, continuing gas buildup.
\end{abstract}

\firstsection 
\section{Introduction}

Cosmic accretion onto galaxies is generally difficult to observe because the
accreting halo gas is mostly ionized and the accretion rate can be irregular. Here
we report fairly clear accretion via two streams of neutral gas onto the local
starburst dwarf Irregular galaxy IC 10 observed with the Jansky Very Large Array
and Green Bank Telescopes (\cite[Ashley et al. 2014]{Ashley14}). We also point out
intriguing starburst hotspots in tadpole-shaped galaxies and extremely low
metallicity dwarf galaxies that have metallicities lower than in the rest of the
galaxy by a factor of $\sim5$ (\cite[Sanchez Almeida et al. 2013]{sanchez13};
\cite[Sanchez Almeida et al. 2014a]{sanchez14a}; \cite[Sanchez Almeida et al.
2015]{sanchez15}). These hotspots suggest fresh accretion and triggering of the
starburst. On the other hand, observations of the far outer parts of 20 dwarf
irregulars show slow star formation with a gas consumption time of 100 Gyr or
more, suggesting no need for cosmic accretion to sustain their activity
(\cite[Elmegreen \& Hunter 2015]{eh15}). Whether normal galaxies accrete at
significant rates or just starbursts do this is an open question. A review of
accretion-fed star formation is in \cite[Sanchez Almeida et al.
(2014b)]{sanchez14b}.

\section{IC 10}

IC 10 is a local group dwarf irregular starburst at a distance of 700 kpc. Neutral
atomic hydrogen has been extensively studied, showing a large peripheral excess
(e.g., \cite[Nidever et al. 2013]{nid13}). A recent analysis of the structure and
motions of this HI suggest accretion along two streams, one from the far side in
the north and another from the near side in the south (\cite[Ashley et al.
2014]{Ashley14}). Figure 1 shows JVLA and GBT maps of the HI velocities. Blue is
approaching and red is receding. If we use the swirling spiral pattern seen in HI
as an indication of the sense of rotation, i.e., so that the spirals are trailing,
then the velocities indicate that the near side of the disk is in the south and
the far side is in the north. Since the approaching velocities are also in the
north, we presume they are on the far side. Thus we see that both the far side and
the near side have streams of HI that are moving toward the center of the galaxy.
This is accretion. There is also an indication of excessive turbulence in the
northeast and southwest, as shown by the velocity dispersion maps (\cite[Ashley et
al. 2014]{Ashley14}). These could be impact sites.

\begin{figure}[b]
\begin{center}
 \includegraphics[width=5in]{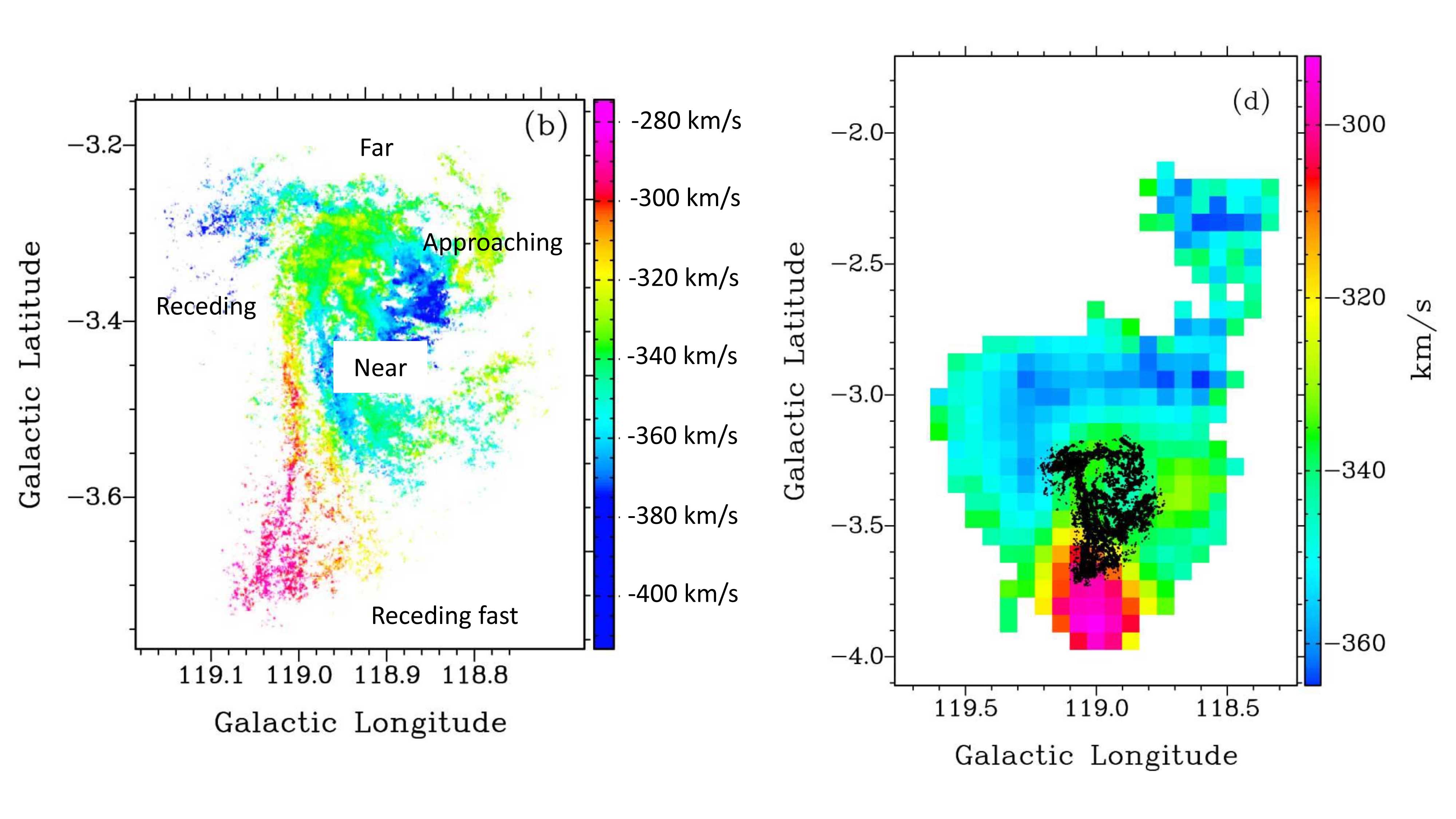}
 \caption{IC10 in HI with the VLA (left) and GBT (right), showing velocities as indicated by the
 color scale. Receding motions (red) in the south and approaching motions (blue) in the north
 indicate accretion toward the galaxy along two streams.  The accretion rate is about equal
 to the star formation rate in this local group starburst. Images
 from \cite[Ashley et al. (2014)]{Ashley14} with labels added.}
   \label{fig1}
\end{center}
\end{figure}

The accretion rates in the northern and southern streams can be estimated from the
velocity gradients and the masses. They are $0.001\;M_\odot$ yr$^{-1}$ in the
north and $0.05\;M_\odot$ yr$^{-1}$ in the south. The star formation rate is
comparable, $0.08\;M_\odot$ yr$^{-1}$.

Clear examples of HI accretion like this are rare and the origin of the streaming
gas is not known. It could be fall-back from a previous tidal interaction, perhaps
with nearby M31, but it is fairly regular in structure and motion and fall-back
would seem to be more cloudy and dispersed. There are no measurements of
metallicity in these accretion streams. We predict it will be lower than in the
main disk of IC 10.

\section{Starburst hotspots with locally low metallicities}

In a study of tadpole-shaped galaxies, which are common at high redshift
(\cite[Elmegreen \& Elmegreen 2010]{ee10}) and among extremely low metallicity
dwarfs locally (\cite[Morales-Luis et al. 2011]{ml11}; \cite[Elmegreen et al.
2012]{e12}), we determined metallicities in the HII regions using several
techniques. In almost all cases (17 out of 23 studied) the metallicities were
significantly lower in the brightest HII regions than elsewhere in the galaxies.
This leads us to suspect that the metal-poor hotspots are accretion regions of
nearly pristine gas.

\begin{figure}[b]
\begin{center}
 \includegraphics[width=5in]{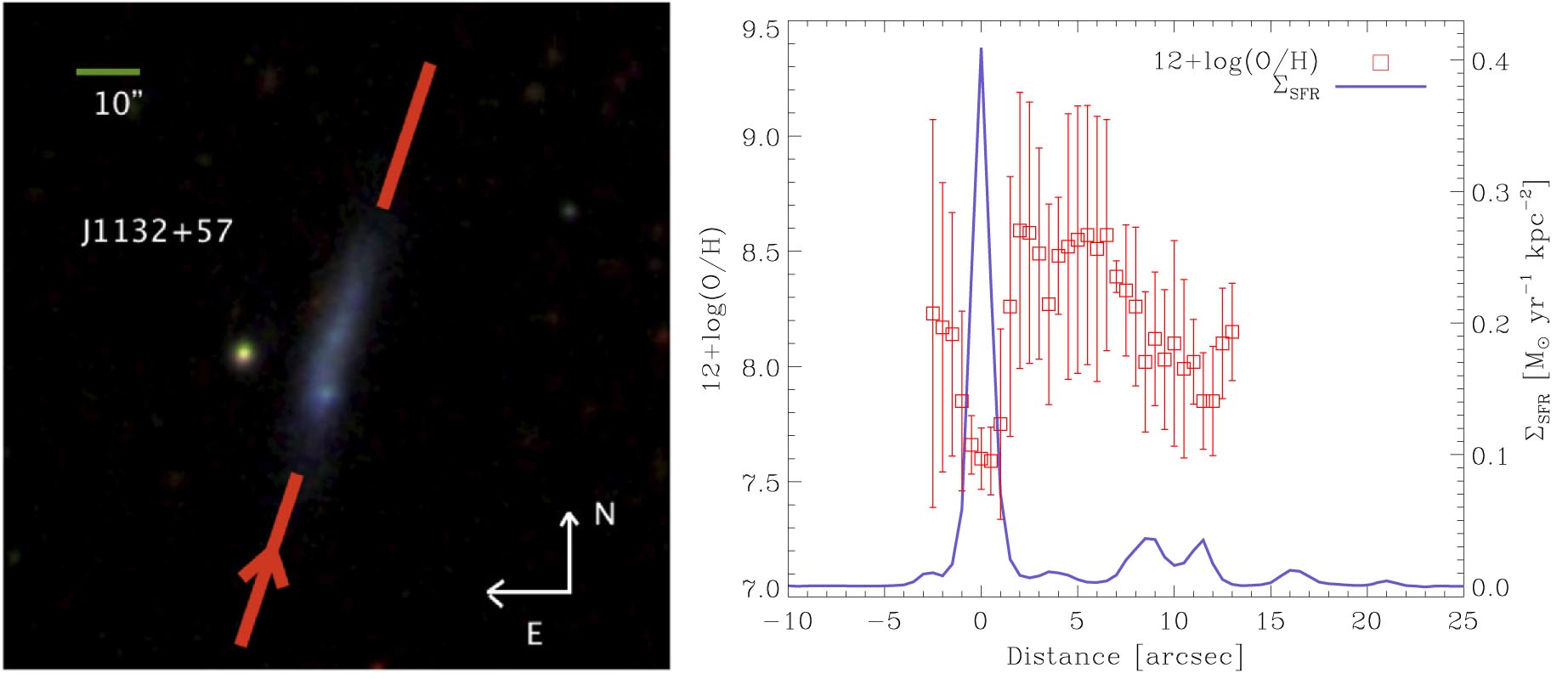}
 \caption{Tadpole galaxy SDSS image (left) with slit position indicated and scans of
 star formation rate (blue) and metallicity (red) on the right. The metallicity drops
 at the site of the peak star formation rate. Images from
 \cite[Sanchez Almeida et al. (2015)]{sanchez15}.}
   \label{fig2}
\end{center}
\end{figure}

Tadpole galaxies are elongated with one bright star-forming region at the ``head''
and a lower brightness tail that may be 5 times larger than the head. Many BCDs
have tadpole or comet shapes (\cite[Loose \& Thuan 1986]{lt86}; \cite[Noeske et
al. 2000]{noeske00}). Presumably most tadpoles are disks viewed edge-on because
they have exponential profiles outside the hotspots and they show rotation
(\cite[Sanchez Almeida et al. 2013]{sanchez13}).

Figure 2 shows an image of a tadpole with a metallicity and intensity trace to the
side. The metallicity was calculated using the method of HII-CHI-mistry13 which
involves a chi-squared fit to many spectral lines (\cite[P\'erez-Montero
2014]{pm14}). The observations were taken with the OSIRIS instrument at the 10-m
Gran Telescopio Canarias.  The metallicity drop at the star formation hotspot is
typical for this type of galaxy.  The smallness of the spot suggests a short
timescale for the accretion, significantly less than a disk orbit time, or
$\sim100$ Myrs. The excess star formation rate corresponds to an excess gas mass
in an accretion event which is consistent with the decrease in metallicity
compared to the rest of the galaxy if the main disk is diluted with nearly
pristine material (\cite[Sanchez Almeida et al. 2015]{sanchez15}). The actual
accretion stream has not been seen in any of these galaxies yet, but it could be
ionized and very faint.

A simulation of an accretion event that could make a tadpole galaxy with a
metal-poor hotspot is in \cite[Verbeke et al. (2014)]{verb14}. Metallicity drops
associated with enhanced star-formation activity driven by gas accretion are also
present in cosmological zoom-in simulations by \cite[Ceverino et al.
(2015)]{cev15}.

\section{Outer Parts of Dwarf Irregulars}

In a recent study of 20 local dwarf Irregular galaxies, \cite[Elmegreen \& Hunter
(2015)]{eh15} measured the properties of HI and star formation to very low
intensity levels. The star formation rate was proportional to the HI gas column
density to a power between 2 and 3, as is typical for dwarfs and the outer parts
of spiral galaxies (\cite[Bigiel et al. 2008]{bigiel08}; \cite[Bigiel et al.
2011]{bigiel2011}; \cite[Kennicutt \& Evans 2012]{ke12}).  The timescale for gas
consumption in the furthest regions where star formation was detected in the FUV
is around 100 Gyr, which is such a long time that no gas is needed to sustain star
formation. The average radial profiles of FUV and HI are also fairly regular and
exponential in shape, which suggests further that there is little active accretion
in the outer parts.

Figure 3 shows the radial profiles of gas surface density (upper left), gas
density (lower left), which is from the ratio of the column density to the scale
height as fit from the stellar and gaseous surface densities; the gas velocity
dispersion (upper right), and the free fall time in the midplane (lower right),
which comes from the midplane density. In each panel, the left-hand axis is the
quantity plotted in natural logs so the exponential scale length can be read from
the figure (each drop by unity in the natural log corresponds to one exponential
scale length on the abscissa). The right-hand axes are the physical units plotted
in a base-10 log.

\begin{figure}[b]
\begin{center}
 \includegraphics[width=5in]{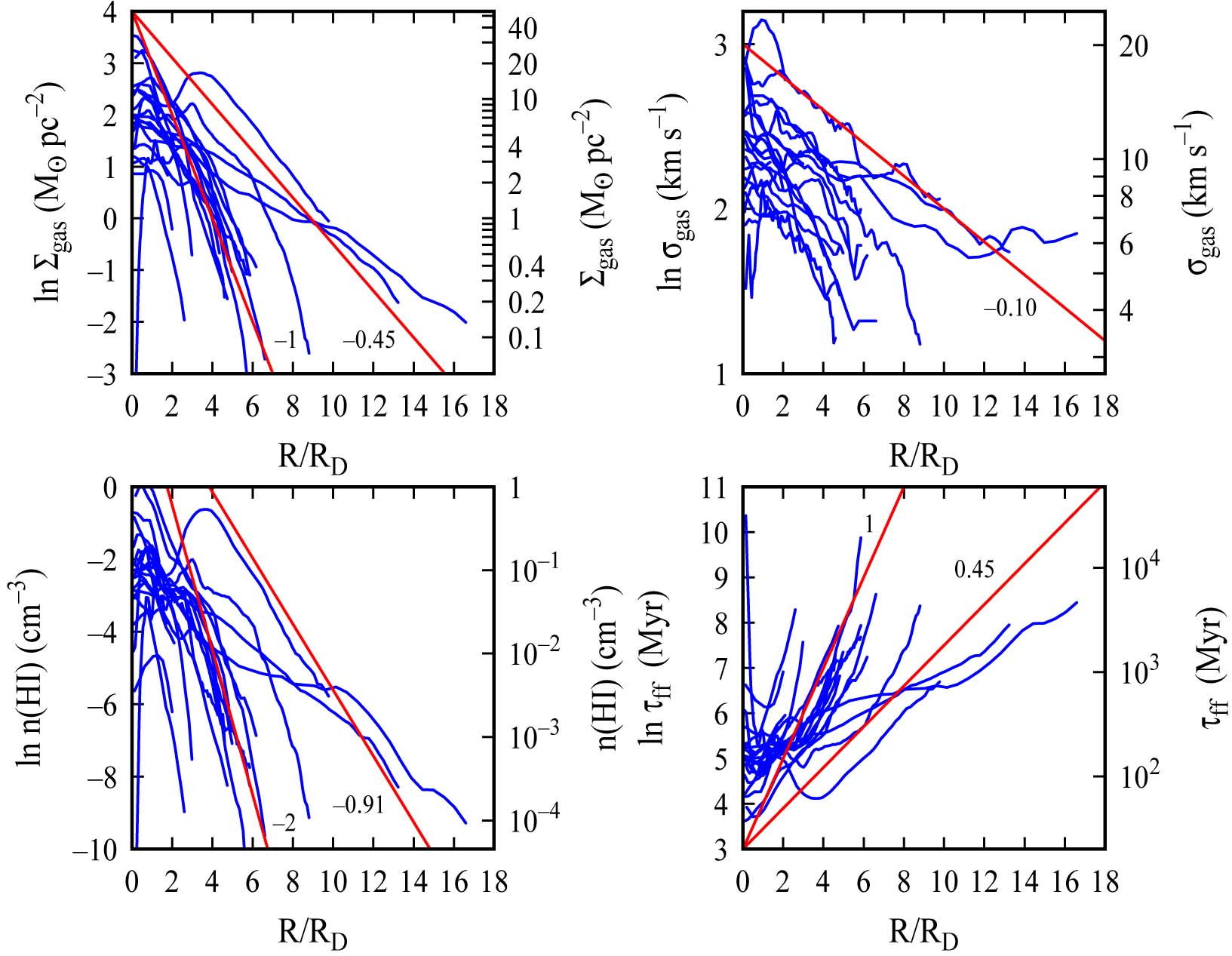}
 \caption{Radial profiles of HI gas surface density (upper left), density (lower left),
 velocity dispersion (upper right), and midplane free fall time (lower right) for 20
 dIrr galaxies showing approximately exponential radial profiles to the far outer regions.
 The free fall time in the outer parts
 corresponds to a gas consumption time of $\sim100$ Gyr, which is so long
 that accretion is not needed to sustain star formation. Image from \cite[Elmegreen \& Hunter
(2015)]{eh15}.}
   \label{fig3}
\end{center}
\end{figure}

The figure shows HI surface densities corrected for projection down to
$\sim0.1\;M_\odot$ pc$^{-2}$, which is one-tenth that of a damped Lyman alpha
cloud. The midplane density drops down to $\sim10^{-3}$ cm$^{-3}$, at which point
the Stromgren radius of an O5 type star is 30 kpc, much larger than the galaxy
(suggesting escape of ionizing photons if there were such a star). The velocity
dispersion continues to decrease with radius, but slowly, and the free fall time
continues to increase with radius, getting to values larger than 1 Gyr.
Considering the $\sim1$\% efficiency of star formation elsewhere in our galaxies,
this free fall time divided by the efficiency is $\sim100$ Gyr, and that is the
gas consumption time, as mentioned above.

\section{Summary}

Three different types of observations about accretion onto galaxies have been
shown. There is evidence for active accretion onto IC 10 in the form of two
in-moving streams of neutral gas with velocity gradients of several km/s/kpc.
These gradients correspond to accretion times on the order of a Gyr. The origin of
this gas is unknown as is its metallicity, but its impact with the disk of this
dwarf irregular could be what is triggering the current starburst, since the
accretion rate equals about the star formation rate.

There is indirect evidence for low-metallicity accretion in a high fraction of
extremely low metallicity galaxies or tadpole-shaped galaxies because the
brightest spots of star formation have metallicities that are lower than in the
rest of the disk by a factor of about 5.  This accretion seems to be in the form
of discrete metal-poor clouds that impact the disk and trigger an excess of star
formation during a $\sim100$ Myr period.

Normal dwarf irregulars seem to have little accretion, however. Their HI is
distributed somewhat uniformly with an exponential radial profile on average, and
their midplane densities are so low that star formation can be sustained at the
present rate for 100 Gyr.  The star-forming disk in FUV is also fairly regular
with an exponential profile.

\end{document}